\documentclass[useAMS,usenatbib]{mn2e}
\usepackage{graphicx}
\usepackage{psfig}
 
\title[A Transition in the Accretion Properties of Radio--Loud
Active Nuclei]{A Transition in the Accretion Properties of Radio--Loud
Active Nuclei}

\author[D. Marchesini, A. Celotti and L. Ferrarese] {Danilo
Marchesini$^1$\thanks{E-mail: danilom@sissa.it (DM); celotti@sissa.it
(AC); lff@physics.rutgers.edu (LF)}, Annalisa
Celotti$^{1}$\footnotemark[1] and Laura
Ferrarese$^{2}$\footnotemark[1]\\ $^{1}$International School for
Advanced Studies (SISSA/ISAS), via Beirut 4, 34014 Trieste, Italy\\
$^{2}$Department of Physics and Astronomy, Rutgers University, 136
Frelinhuysen Road, Piscataway, NJ 08854, U.S.A.}

\begin{document}



\maketitle


\begin{abstract}
We present evidence for the presence of a transition in the accretion
properties of radio-loud sources. For a sample of radio galaxies and
radio--loud quasars, selected based on their extended radio properties, the
accretion rate is estimated from the black hole mass and nuclear
luminosity. The inferred distribution is bimodal, with paucity of sources
at accretion rates, in Eddington units, of order $\sim 10^{-2}$ - assuming
a radiative efficiency of 10 per cent - and possibly spanning one-two
orders of magnitude.  Selection biases are unlikely to be responsible for
such behavior; we discuss possible physical explanations, including a fast
transition to low accretion rates, a change in the accretion mode/actual
accretion rate/radiative efficiency, the lack of stable disc solutions at
intermediate accretion rates or the inefficiency of the jet formation
processes in geometrically thin flows. This transition might be analogous
to spectral states (and jet) transitions in black hole binary systems.
\end{abstract}


\section{Introduction}

Radio-loud active galactic nuclei (AGNs) are characterized by the
presence of relativistic jets, whose energetics at least in some
sources is comparable to that produced by the nuclear engine.  Why and
how some systems are able to form powerful jets remains one of the
unsolved issues in AGN research, and currently limits our
understanding of the relation between the active nucleus, its host
galaxy and the larger scale environment. It is possible that all AGNs
pass through a jet phase (evolutionary interpretation) or that only
certain objects develop powerful jets. For further progress to be
made, it is crucial to determine if and how the nuclear activity, in
terms of accretion properties onto the central black hole (BH) and/or
the black hole spin, is related to the presence and formation of
jets. MHD processes appear to be most likely involved in the
extraction (and collimation) of energy to form jets from the accretion
disc \citep{blandpay82} and/or black hole spin \citep{blandzna77}. A
significant amount of theoretical work has been carried out recently
in order to understand the effectiveness of these processes -- which
is largely related to the strength and topology of the magnetic field
-- in relation to the properties/regime of accretion
(e.g. \citealt{ghosh97}; \citealt{livio99}; \citealt{balbushawley02};
\citealt{meier02}; \citealt{merloni02}; \citealt{livio03}).

A further relevant element tightly related to this issue is the evidence
that jets appear in two rather neatly distinct flavors, the more powerful
ones exhibiting FR~II radio morphology, the weaker ones FR~I morphology
\citep{fanarof74}. At the zeroth order, the unified model for radio--loud
AGNs identifies these two typologies of radio galaxies (RGs) with the
parent populations (i.e. the misaligned counterparts with respect to the
jet direction) of radio-loud quasars (RLQs) and BL~Lac objects
respectively, whose observed properties are dominated by the highly
anisotropic emission from relativistic emitting jets \citep{blandrees78}.
This association is supported by the similar isotropic properties as well
as the relative number density of such (parent and beamed) populations
(\citealt{urry95} for a review).  For the high power systems the lack of
broad emission lines in FR~II RGs has been interpreted as due to the
presence of material obscuring the line of sight at large angles with
respect to the jet axis (torus).  The existence of an obscuring torus has
found direct support from the detection of scattered broad emission lines
and from infrared studies which have revealed quasar nuclei hidden by rest
frame visual extinction in several FR~IIs (e.g.  \citealt{antonucci84};
\citealt{hill96}).

Although this scenario is successful at the zeroth order, issues are
still open.  On a rather robust observational ground, it has emerged
that the fraction of broad lined objects increases with the source
luminosity, at least at low luminosities, and thus - in flux limited
samples - with redshift (however this appears not to be the case when
weak narrow line objects are excluded,
\citealt{willott01}). Furthermore the quasar narrow line ([OIII])
luminosity and submillimeter emission are larger than those of radio
galaxies (e.g. \citealt{willott02a} for a recent overview).  The two
main interpretations of such findings invoke either a dependence of
the torus opening angle on source luminosity (`receding torus model',
\citealt{lawrence91}; \citealt{hill96}) or the presence of a
population of - low luminosity - FR~IIs with low levels of nuclear
accretion emission (\citealt{hine79}; \citealt{laing94};
\citealt{baum95}; \citealt{jackson99}; \citealt{willott00};
\citealt{chiab}; \citealt{grimes03}). Another puzzling aspect is the
recent suggestion that there might exist a so far undetected
population of QSOs with FR~I radio morphology \citep{blundrawl01}.

Furthermore, despite of the clue offered by the findings by
\citet{ledlow96} that the radio power at the transition between FR~Is
and FR~IIs is correlated to the host galaxy magnitude, it is still
unclear what causes the FR~I-FR~II dichotomy. Possibilities include
the different ways in which a jet interacts with the ambient medium
(\citealt{bick95}; \citealt{gkwi01}) and/or the different nuclear
intrinsic properties of the accretion and jet formation and the jet
content (\citealt{baum95}; \citealt{meier99}; \citealt{rey96}).  While
little observational evidence has been gathered in support of any of
these interpretations, the use of correlations between the global
large scale and nuclear properties for the sample analyzed by
\citet{ledlow96} intriguingly suggests that a fundamental parameter
determining the FR~I-FR~II dichotomy could be the mass accretion rate
in Eddington units, which possibly regulates the accretion mode and
also the characteristics of nuclear outflows
(\citealt{ghisellinicelotti2001}).  Therefore, while the environment
could play a role in determining the radio morphology and the jet
appearance, it is of fundamental importance to determine any
connection between the accretion properties and formation of jets.

In this context the aim of our work is to shed light on the process of
formation and evolution of jets with respect to the nuclear AGN
properties. In particular we intend to assess the role of the central
black hole mass ($M_{\rm BH}$) and the mass accretion rate (in
Eddington units, $\dot{m}\equiv \dot M / \dot M_{\rm Edd} $) in
determining the nuclear and extended properties for a sample of
radio-loud objects (RGs and RLQs) as complete as possible.  Clearly,
$\dot{m}$ relies on an estimate of $\dot{M}$, which will be
(necessarily) inferred from the bolometric accretion luminosity of the
AGN, $L_{\rm bol}$. The unknown radiative efficiency of the accreting
flow will constitute a crucial parameter in interpreting our findings.

The outline of the paper is the following.  In Section~2 we introduce
the considered samples of sources; in Section~3 we discuss the
analysis and assumptions adopted to obtain the central black hole
mass, the bolometric luminosity and the mass accretion rate in
Eddington units.  The results and their discussion are presented in
Section~4 and 5, respectively, while in Section~6 we summarize our
conclusions. For comparison with previous work in this field a zero
cosmological constant universe with $H_{0}$=75 km s$^{-1}$ Mpc$^{-1}$
and $q_{0}$=0.5 is adopted [Assuming a $\Lambda$CDM cosmology
($\Omega_{\Lambda} =0.7, \Omega_{\rm M} = 0.3$) does not alter our
results, and quantities can be easily re-estimated from Table~1].

\section[]{Sample selection}

Our goal is to investigate the nuclear properties of a well defined
sample of radio--loud objects, whose selection is unbiased with
respect to the nuclear properties themselves. Our starting point are
the 298 RGs and the 53 RLQs from the 3CR catalogue \citep{spinrad85}.
 
For each source, we must be able to estimate $M_{\rm BH}$ (from the
stellar luminosity of the host galaxy, $M_{\rm bulge}$, see \S 3.1),
as well as the bolometric luminosity of the active nucleus.  While
$M_{\rm bulge}$ is easily measured for RGs, in RLQs the stellar light
is often overwhelmed by the non-thermal contribution of the AGN. From
the complete sample of 53 3CR RLQs we have therefore selected the
sub-sample for which the host galaxy is clearly resolved
\citep{lehnert99}.  This includes 31 objects with 0.3$\la z \la$2.0
and total radio luminosities at 178~MHz in the range 10$^{34}$ $<
L_{178}<$ 10$^{36}$ erg~s$^{-1}$~Hz$^{-1}$.\footnotemark
\footnotetext{Twelve of the 3CR RLQs were not observed by Lehnert et
al. because of scheduling problems.}

\begin{figure}
\center \includegraphics[width=8cm]{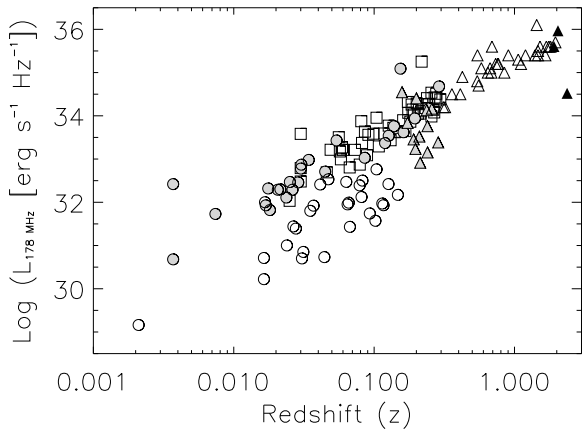}
\caption{Total radio luminosity $L_{178}$ at 178~MHz versus
redshift. Circles are FR~Is (empty ones are B2RGs while filled ones are
3CR~RGs), squares are FR~IIs and triangles are RLQs (empty ones are
3CR~RLQs; light filled ones are MDRLQs and the dark filled ones are
the three RLQs from \citealt{hutch2002}).}
\label{fig4}
\end{figure}

On the other hand, while $L_{\rm bol}$ is relatively easy to estimate
(modulo an appropriate bolometric correction) for RLQs, as the
emission from the central AGN is bright and easily measurable, the
situation is more delicate for RGs. To obtain the best possible
constraints on $L_{\rm bol}$, we select the 79 3CR radio galaxies for
which high resolution Hubble Space Telescope (HST) images led to
either a measurement or an upper limit of the optical unresolved
nuclear luminosity\footnote{3CR~386 was included in the Chiaberge et
al.'s study. However, ground based observations subsequently confirmed
the putative nucleus to be a foreground star (Marchesini, Ferrarese \&
Celotti, in prep). We will exclude the galaxy from further
consideration.}  (Chiaberge, Capetti \& Celotti 1999, 2000, 2002).
This sample includes 80\% of the 3CR RGs with $z \leq 0.3$. Of these,
26 are FR~Is (10$^{31}$ $< L_{178}<$ 10$^{35}$ erg~s$^{-1}$~Hz$^{-1}$)
and 53 FR~IIs (10$^{32}$ $< L_{178}<$ 10$^{35}$
erg~s$^{-1}$~Hz$^{-1}$).  While not complete, this sample is
homogeneous (all galaxies belong to the radio-flux limited 3CR sample)
and unbiased with respect to nuclear properties (see discussion in
\S5).  We will consider this our main sample. We also note the
significant overlapping of the sources both in $L_{178}$ and $z$ (see
Fig.~\ref{fig4}).

In order to assess the incidence of possible selection biases, enlarge
the statistics, and include RLQs with redshifts overlapping with the
redshift range of the RGs (thus reducing the redshift-luminosity
degeneracy and any evolutionary effect, see Fig.~\ref{fig4}), we
also consider three additional samples. The first comprises the 28 low
luminosity RGs (mostly FR~I) from the B2 sample (B2RG) with published
high resolution HST images \citep{B2_02}. These galaxies populate the
low redshift range $z < 0.15$ and span 10$^{29}$ $< L_{178}<$
10$^{33}$ erg~s$^{-1}$~Hz$^{-1}$. The second sample includes the 12
RLQs from \citet{McLureDunlop2001} (MDRLQ). This is a rather
heterogeneous sample with redshift 0.15$\la z \la$0.3; the common
thread being the availability of HST images for all objects. The third
sample includes three high redshift (1.9$\la z \la$2.4) RLQs imaged
with HST by \citet{hutch2002}.

The total radio luminosity-redshift plane for all of the sources
considered in this work is shown in Fig.~\ref{fig4}.

\section{Analysis}

In this section we describe the steps taken to estimate $M_{\rm BH}$,
the AGN bolometric luminosity, and the associated mass accretion rate
$\dot{M}=L_{\rm bol}/\epsilon c^{2}$, where $\epsilon$ is the
radiative efficiency.

\subsection{Black hole masses}

Only a handful of nearby quiescent galaxies, and only four of the RGs
and RLQs in our sample, namely 3CR~270 (NGC~4261,
\citealt{ferrarese96}), 3CR~272.1 (NGC~4374, \citealt{boweretal98}),
3CR~274 (M~87, \citealt{macchetto97}) and B2~2116+26 (NGC~7052,
\citealt{vanderMarel98}), have $M_{\rm BH}$ measured directly through
dynamical modeling of stellar and gas kinematics. For these galaxies,
$M_{\rm BH}$ correlates tightly with the stellar velocity dispersion
of the host bulge and (with significantly larger scatter) with the
host bulge magnitude (\citealt{kormendy95}; \citealt{ferrarese00};
\citealt{gebhardt00}). Both correlations have been shown to hold for
nearby AGNs (\citealt{ferrarese01}; \citealt{McLureDunlop2002}), and
we will make the fundamental assumption that they can be extended to
the RGs and RLQs considered in this paper. In particular, since
measurements of the stellar velocity dispersion are available for only
18 of our objects (of which only four are recent measurements), we
will employ the following characterization of the $M_{\rm BH} - M_{\rm
bulge}$ relation:
\begin{equation}\label{eq1}
\log{M_{\rm BH}} = -1.58(\pm 2.09) -0.488(\pm 0.102)M_{\rm B},
\end{equation}
where $M_{\rm BH}$ is in solar masses and $M_{\rm B}$ is the total
$B-$band luminosity of the host elliptical galaxy. This is derived for
the most current sample of 16 elliptical galaxies for which a high
quality measurement of the mass of the central black hole exists
(\citealt{gebhardt03}, and references in \citealt{merritt01}).  The
reduced $\chi^2$ of the relation is 4.7, significantly smaller than
the one ($\chi^2 = 23$) reported for the complete sample of galaxies,
including lenticulars and spirals, with measured $M_{\rm BH}$
\citep{ferrarese00}.  Because of worries about the reliability of
$M_{\rm bulge}$ measurements in spirals, and since all of the galaxies
considered in this paper are genuine ellipticals, we will use eq.~1
for our analysis.  The scatter in eq.~1 corresponds to a 42 per cent
uncertainty in $M_{\rm BH}$, slightly larger than the value (30 per
cent) quoted by \citet{McLureDunlop2002} based on a now out-of-date
sample.

For the 3CR sources, rest-frame $V-$band absolute magnitudes are taken
from the references listed in \citet{zirbaum95} or, when not
available, from the apparent $V-$band magnitudes reported in the 3CR
catalogue.  Corrections for Galactic extinction were estimated mostly
from Burstein \& Heiles (1982, 1984) and - when not possible - from
\citet{schlegel98}. In all cases, a color correction $B-V$=0.96,
appropriate for elliptical galaxies, was applied \citep{fukugita95}.
For the galaxies in the B2 sample, $M_{\rm B}$ is from
\citet{impey93}. $R$-band magnitudes of the host galaxies of the 3CR
RLQs have been estimated from HST images \citep{lehnert99} from the
residual luminosity after subtraction of the nuclear PSF, within an
aperture of radius 1$^{\prime \prime}$.4 (which, at the redshift of
the targets, covers the entire galaxy).  For the RLQs in the McLure \&
Dunlop and Hutchings et al. samples, $M_{\rm R}$ are listed by the
authors. A color correction $B-R$=1.57 (Fukugita et al. 1995) was
applied to transform $R-$ to $B-$band magnitudes.  Cosmological and K
corrections have been applied to all magnitudes.

For each galaxy, the rest-frame absolute magnitudes and the BH masses
(from eq. 1) are listed in Table~1.

\subsection{Bolometric luminosities}

All of the objects in our sample have HST images.  With the exception
of the RLQ sample of McLure \& Dunlop, the optical nuclear luminosity,
$L_{\rm o}$, is derived by fitting the unresolved nuclear component
with the appropriate HST point spread function (Chiaberge et al. 1999,
2002; \citealt{B2_02}; \citealt{hutch2002}).  For the RLQ sample of
McLure \& Dunlop, $L_{\rm o}$ is obtained from the rest-frame
monochromatic luminosity at 5100 \AA~ \citep{McLureDunlop2001}.
K-corrections have been applied in all cases.  Uncertainties on
$L_{\rm o}$ vary for the different samples, but are typically of the
order of 15 per cent.

Estimating the nuclear bolometric luminosity $L_{\rm bol}$ from
$L_{\rm o}$ involves the knowledge of the bolometric correction,
$\eta$, which we take from \citet{elvis94}. This was derived for a
sample of radio-loud and radio-quiet quasars, and might not be
appropriate for the radio galaxies considered here. We will discuss
the possible consequences of our assumptions in \S 5.1.

Nuclear bolometric luminosities are listed in Table~1 for all objects.
The uncertainty on the adopted $\eta$ is $\sim$35 per cent, yielding
an average error for $L_{\rm bol}$ of $\sim$ 38 per cent.

\section{Results}

In the following, we will discuss whether different classes of objects
in our sample have different properties in terms of BH mass, nuclear
bolometric luminosity and/or mass accretion rate. For radio galaxies,
we will make the distinction between FR~I and FR~II. Furthermore, in
view of the evidence that the FR~II population might be heterogeneous
in terms of its nuclear properties (see \S 5) the FR~II RGs will be
further divided according to their optical spectral properties in low-
and high-ionization emission line galaxies (LEGs and HEGs,
respectively) and broad line RGs (BLRGs).

\subsection{Black hole masses}

\begin{figure}
\center \includegraphics[width=7.5cm]{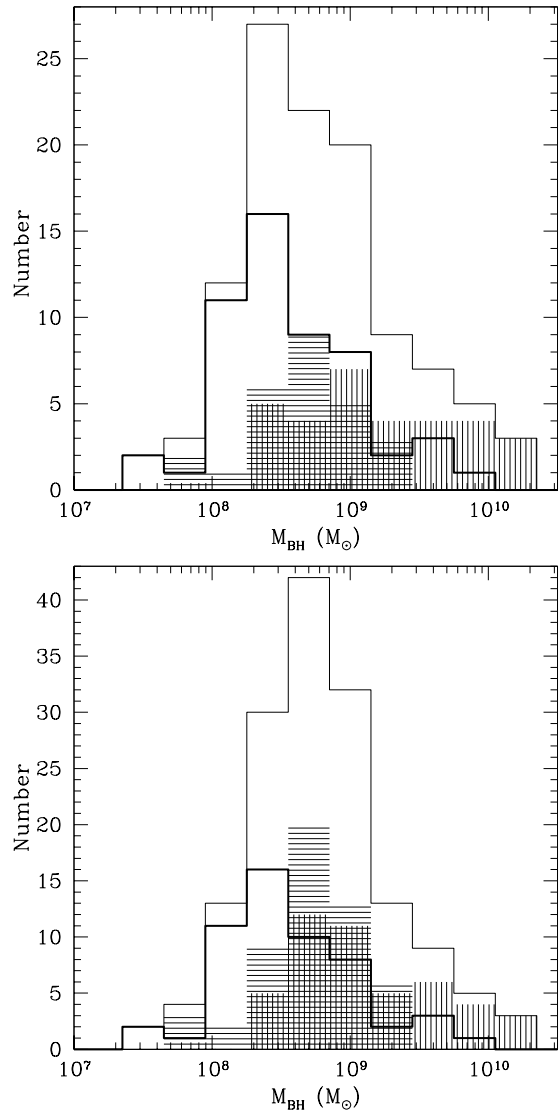}
\caption{Distributions of $M_{\rm BH}$ for the 3CR sources (upper
  panel) and for all of the sources considered in this work (lower
  panel): the thin continuous line is the distribution for all of the
  objects, the horizontal shaded area represents the FR~Is, the
  vertical one represents the distribution for the RLQs and the thick
  continuous line indicates the FR~IIs.}
\label{fig1}
\end{figure}

Although the objects in our sample span about three decades in $M_{\rm
BH}$ (Fig.~\ref{fig1}), most BH masses are in the
$10^8-10^9$M$_{\odot}$ range. The paucity of objects (at least radio
galaxies) hosting BH with smaller mass is possibly due to selection
effects (although the samples were selected based on radio power, see
\citealt{McLureDunlop2002}; but also \citealt{ho2002};
\citealt{oshlack2002}; \citealt{woo02a}).

The upper panel in Fig.~\ref{fig1} shows the distribution of black hole
masses $M_{\rm BH}$ for all 3CR~sources (FR~I, FR~II and RLQ); the lower
panel shows the distribution of $M_{\rm BH}$ for all of the objects
considered in this study.

A Kolmogorov-Smirnov statistical test (Table 2) shows significant
differences in the distribution of $M_{\rm BH}$ only between RLQs and
both FR~Is and FR~IIs, but with mean values within their standard
deviations.  No evidence for differences are found for objects
belonging to different sub-classes or different samples (3CR RGs vs B2
RGs, or 3CR RLQs vs RLQs from McLure \& Dunlop).
 
\subsection{Bolometric luminosities}

\begin{figure}
\center
\includegraphics[width=7.0cm]{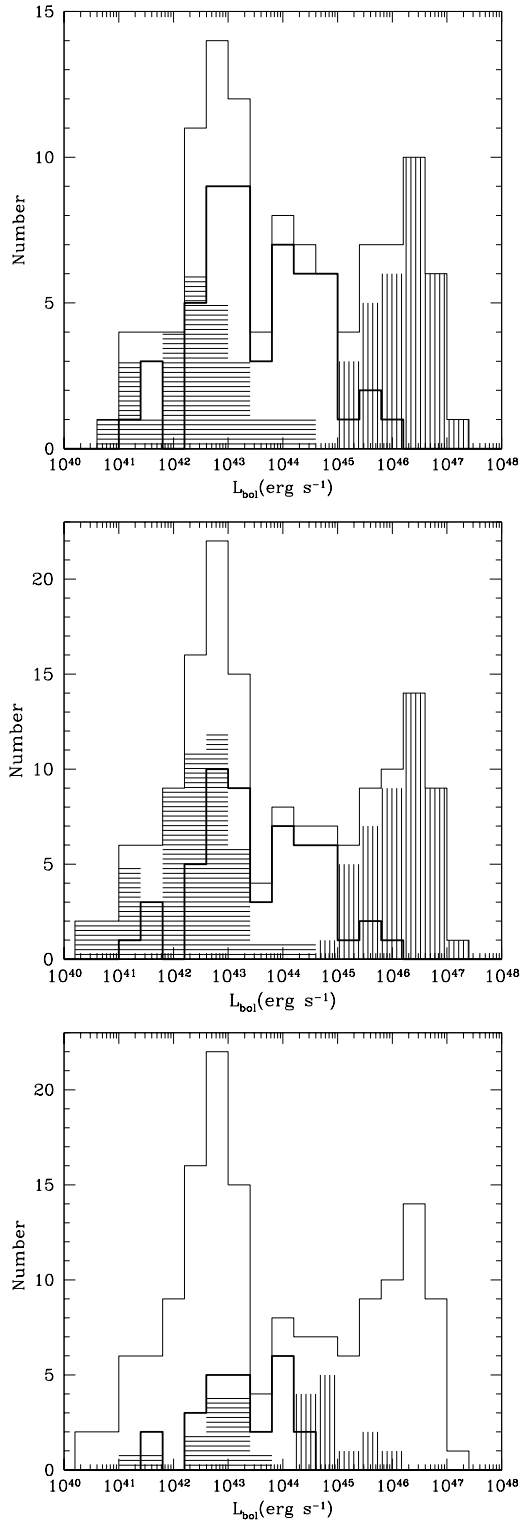}
\caption{Distributions of $L_{\rm bol}$ for the FR~Is, FR~IIs and RLQs
sources belonging to the 3CR~sample (upper panel) and for all of the
objects considered in this work (middle panel): the thin continuous
line is the distribution for all of the objects; the horizontal shaded
area represents the FR~Is, the vertical one represents the RLQs and
the thick line refers to the FR~IIs. The lower panel shows the
distributions of $L_{\rm bol}$ for the LEG (horizontal shaded area),
HEG (thick continuous line) and BLRG (vertical shaded area)
sub-samples; the thin continuous line is the distribution for all of
the objects.}
\label{fig2}
\end{figure}

Fig.~\ref{fig2} shows the distribution of nuclear bolometric
luminosity $L_{\rm bol}$ for FR~Is, FR~IIs and RLQs in the 3CR sample
(upper panel) and for all of the sources in this study (middle panel).
Unlike the distribution in $M_{\rm BH}$, objects belonging to
different classes show systematic differences in their nuclear
(optical or bolometric) luminosity (see Table 2). This is neither a
new nor an unexpected result (indeed the distinction between RGs and
RLQs is based on their nuclear luminosity, see also \S 5). However, it
is intriguing to notice that the overall distribution - spanning about
seven orders of magnitude - appears structured, with two peaks easily
recognizable, and a shortage of sources at $L_{\rm bol} \sim 10^{44} -
10^{45}$ erg s$^{-1}$.  The low luminosity peak ($L_{\rm bol} \sim
10^{42.7}$ erg s$^{-1}$) is comprised in equal measure of FR~I and
FR~II sources, while the high luminosity peak includes mainly
radio-loud quasars. While there are no FR~I galaxies straddling the
two peaks, some FR~IIs extend to the high luminosity regime populated
by the quasars.

We estimated the statistical significance of a possible bimodality
using the KMM algorithm \citep{kmm94}. The distributions for the
entire sample, and for the 3CR sample alone (both RGs and RLQs) are
both strongly inconsistent with being unimodal (P-value $<$0.05, see
Table 2). The distribution of the 3CR radio galaxies alone,
however, is consistent with unimodality (P-value $>$0.1),
although there is marginal evidence for a second peak around 10$^{44}$
erg s$^{-1}$. This peak is populated exclusively by FR~II sources; if
these are further divided into LEGs, HEGs and BLRGs, an additional
trend emerges: all LEGs and about 2/3 of the HEGs are low luminosity
sources, while 1/3 of the HEGs and all of the BLRGs have high nuclear
luminosity (see Fig.~\ref{fig2} lower panel).  We will discuss this
point in more detail in \S 5. We point out that sources for which only
upper limits to the nuclear luminosity exist, are found exclusively in
the low luminosity peak (see Table 2), thus not affecting these
findings (and the equivalent ones about the $\dot m$ distributions).

\subsection{Mass accretion rates}

\begin{figure}
\center \includegraphics[width=7.0cm]{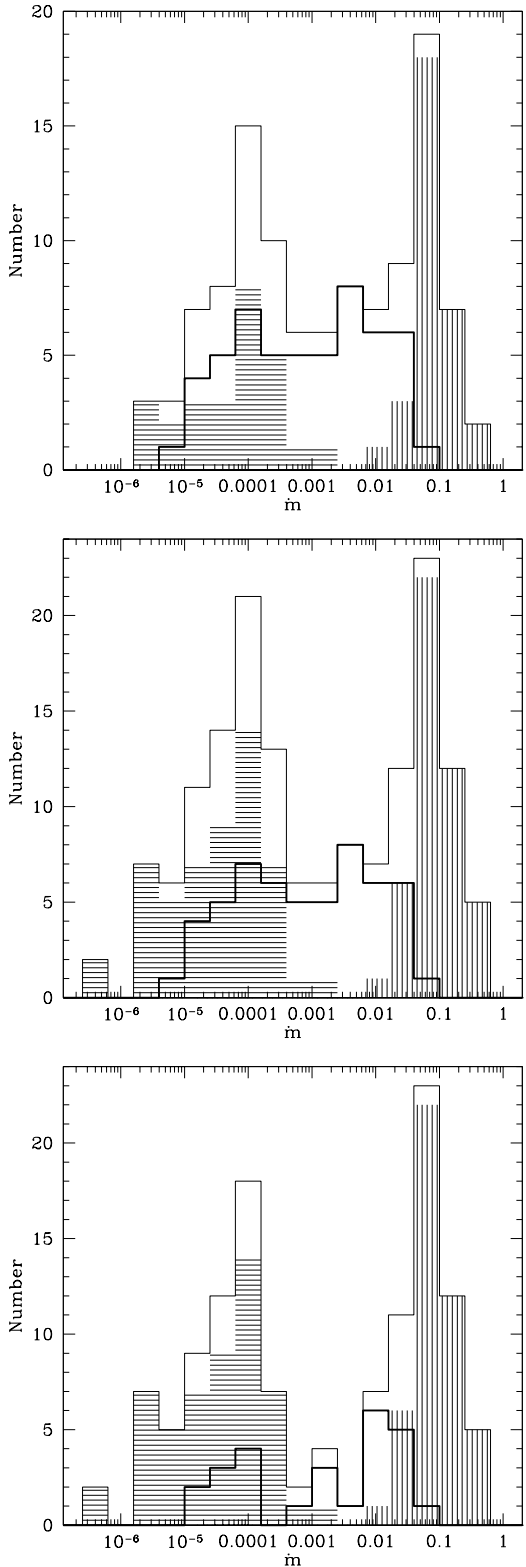}
\caption{Distributions of $\dot{m}$ for the FR~Is, FR~IIs and RLQs
belonging to the 3CR~sample (upper panel), for all of the sources
studied in this work (middle panel): the thin continuous line
represents all the sources, the horizontal shaded area indicates the
FR~Is, the vertical shaded area the RLQs and the thick line represents
the distribution for the FR~IIs. The lower panel is the same as the
middle one, but only LEGs and BLRGs are represented by the thick
continuous line, while HEGs are excluded. The reported values
correspond to a radiative efficiency $\epsilon = 1$.}
\label{fig3}
\end{figure}

The mass accretion rate in Eddington units is then easily estimated as
$\dot{m} = L_{\rm bol}/\epsilon L_{\rm Edd}$. The resulting
distribution is shown in Fig.~\ref{fig3} for the 3CR~sample (upper
panel) and for all the considered objects (middle panel), under the
assumption (to which we will return in \S 5) that the radiative
efficiency $\epsilon$ is equal to unity for all sources.

A KMM test (Table 2) shows that the overall distribution in $\dot m$
is non unimodal (P-value $<$0.05), a clear consequence of the
fact that, while the distribution in $M_{\rm BH}$ is unimodal, the one
in $L_{\rm bol}$ is not.  In this case, the $\dot m$ distribution for
the 3CR radio galaxies alone is also inconsistent with being unimodal,
(P-value $<$0.05; see Table~2), although the $L_{\rm bol}$
distribution for the same sample is consistent with being unimodal.
Note that the typical uncertainty on $\dot m$ ($\Delta \log{\dot{m}}
\approx$0.45) is smaller than the width of the peaks and significantly
smaller than their separation. All of the BLRGs and 1/3 of the HEGs
have $\dot{m} \ga$2$\times$10$^{-3}$, while all LEGs and 2/3 of the
HEGs have $\dot{m} \la$2$\times$10$^{-3}$. The LEGs have statistically
the same distribution in $\dot{m}$ as the FR~I galaxies (see Table~2),
while HEGs have an asymmetric distribution, peaking at $\dot{m}
\sim$3$\times$10$^{-4}$ and rapidly declining toward smaller
$\dot{m}$.

As a final test on the presence of bimodality we also considered the
effect of the dispersion in the $M_{\rm BH}-M_{\rm B}$ correlation
from which we estimated the BH masses. By adopting a Gaussian
probability distribution of $M_{\rm BH}$ (for a given $M_{\rm B}$)
with width equal to the scatter of such relation we found that the
bimodality is still significant, but only when all of the 3CR sample
(and not only the radio galaxies) is considered.

\section{Discussion}

The main findings of this work are: a) within the sample considered,
FR~Is, FR~IIs and radio-loud quasars are indistinguishable based on
$M_{\rm BH}$ which power the nuclear activity, however b) they not
only differ significantly in both nuclear bolometric luminosity and
mass accretion rate (as several previous works have already pointed
out), but there is evidence for a bimodal behavior in such
properties. We stress that the errors associated with $L_{\rm bol}$
and $M_{\rm BH}$ translate, as mentioned before, to an uncertainty on
the mass accretion rate smaller than the width of the peaks.  It is
conceivable that samples with smaller uncertainties on $M_{\rm BH}$
and $L_{\rm bol}$ would produce an even more pronounced multi-modal
distribution in $\dot{m}$ than the one shown in Fig.~\ref{fig3}.

As shown by Chiaberge et al. (2002), the nuclear optical core
luminosity appears to be markedly different for FR~I and most of FR~II
radio galaxies. The low FR~I optical luminosity well correlates with
the radio one, suggesting that the emission in the two bands has a
similar origin, namely non-thermal radiation from the jet.
Furthermore, the high fraction of detected nuclear sources imply the
absence of a geometrically thick obscuring torus, as also supported by
the lower level of mid and far IR emission and limited UV
absorption in FRIs (\citealt{heck}; \citealt{chiab}). FR~II sources
show instead a more complex behavior, as expected given the already
proposed view - based on the optical spectral properties - that the
FR~II class is not a homogeneous population \citep{laing94}.  Some
FR~IIs follow the same correlation as FR~Is, while others show a
significant `excess' in their optical luminosity, given their radio
power. This excess has been interpreted as the contribution from a
thermal component associated with radiation from accretion (see also
the results by \citealt{willott02b} on the anti-correlation between
the 4000 \AA~Ca break and the line luminosity).  This hypothesis is
strongly supported by the fact that the different dominance of
non-thermal vs thermal contribution correlates with the optical line
properties: LEGs behave like non-thermal FR~Is, while BLRGs show a
thermal excess. HEGs show both behaviors. As suggested by the large
line equivalent width of HEGs with apparent FR~I--like optical
luminosity, and in agreement with the basic unification scenario, this
is consistent with HEGs being thermal BLRG-like objects whose optical
emission is partly obscured (Chiaberge et al. 2002).  In this
scenario, low-ionization FR~II and FR~I sources belong to the same
population, while high-ionization FR~IIs and BLRGs correspond to
misaligned radio-loud quasars \citep{jackson99}.  This unification
scheme also accounts for the findings that some BL~Lacs have both
extended radio morphology \citep{kollgaard92} and extended radio power
\citep{cassaro99} typical of FR~II sources, as well as for the
behavior of the luminosity function and evolution of radio galaxies
(e.g. \citealt{dunlpeak}; \citealt{willott99}; Grimes et al. 2003).

How do our findings fit into this overall picture?  In the above scenario,
the dichotomy between non-thermal (jet) and thermal (accretion) dominated
sources is reflected into the observed bimodal distribution in $L_{\rm
bol}$. This implies that all of the HEGs 
would shift toward higher $L_{\rm bol}$ if their intrinsic nuclear
luminosity could be observed: although the amount of absorption is not
known a priori, the above unification scenario implies that the HEG
intrinsic luminosity is comparable to that of BLRGs/RLQs (as appears
to be the case for the couple of sources with spectropolarimetric
data, Chiaberge et al. 2002, see also \S5.1).  Furthermore, because
the observed optical luminosity of FR~Is and LEGs is dominated by the
non-thermal component, it formally represents an upper limit to the
thermal accretion luminosity of these sources. If the two
contributions could be separated, and only the thermal accretion
optical luminosity were plotted in Fig.~\ref{fig2}, all low luminosity
sources would shift toward even lower $L_{\rm bol}$\footnote{Note also
that some of the nuclear luminosities in the low-luminosity range
already represent upper limits, see Table 1.} (see however below),
thus exacerbating the 'separation' between the two populations of
thermal and non-thermal dominated objects.

Before proceeding any further, it is essential to analyze possible
observational biases which might be responsible for the observed
behavior. We begin by pointing out that only two of the 31 3CR RLQs
are flat spectrum. Although for these objects $L_{\rm bol}$ might be
contaminated by non-thermal emission, removing them would not affect
our findings. Furthermore, observational biases are expected to play
against the brightest and more distant RLQs, rather than the weaker
ones, since resolving the underlying galaxy was a necessary condition
for selecting these objects.  It is therefore unlikely that our sample
is missing low luminosity RLQs which would populate the transition
regime in either $L_{\rm bol}$ or $\dot m$. On the other hand, no bias
is expected against bright optical radio galaxies in the complete 3CR
sample observed with HST, since the selection of these sources is
based on the extended radio power, which does not correlate well with
the nuclear optical luminosity. Furthermore, the distribution of the
thermal and non-thermal sources in terms of radio power is continuous:
the two classes overlap, and show no sign of a `gap'.

The observed bimodality in the distribution of $\dot m$ should also
not be induced by possible cosmic evolution effects for the following
reasons. First, the distribution in $\dot m$ is bimodal also when the
sample of 3CR RGs alone is considered; since these sources are all
within $z<$0.3, and since the FR~I and the FR~II sub-samples overlap
both in redshift and extended radio luminosity distributions,
evolutionary effects (if assumed to be similar to those affecting
RLQs) are not crucial for the 3CR RG sample. Second, the lower and the
higher redshift samples of RLQs have similar properties ($\dot
m$). Finally, even the cosmic evolution of the high-z 3CR RLQs,
estimated by using the parameterization $L_{\rm opt} \propto (1+z)^3$
\citep{boyleterlevich98} between $z=1$ and $z=0.1$, is a factor
$\sim$6 (and $\sim$20 for the few $z\sim 2$ sources), much smaller
than the separation between the two peaks of the distribution in $\dot
m$ and therefore not enough for being responsible for the observed
bimodality.

A further possible source of bias could be present if -- as suggested
by \citet{lawrence91} (see also Simpson 1998, 2003) -- there is an
inverse dependence between the degree of obscuration and the source
luminosity (`receding torus model') within the FRII/RLQ
population. This would imply a natural decrease in the number of
detectable sources at lower $\dot m$ and on average a lower luminosity
for the obscured objects.  While there is some observational support
to this scenario (see the above papers), there are some indications
that the bias thus introduced would not alter our results.  We first
notice that our sample is selected on the basis of the extended radio
properties, and while indeed some of the sources have not been looked
at by HST, this was not related to their optical nuclear properties --
in fact the unobserved objects include both HEGs, LEGs and BLRGs in
similar number (only a few for each sub--class).  Secondly, as
discussed below, the bolometric luminosity estimated for HEGs from
their line emission is compatible with that of BLRGs, as also
supported by inferences for objects with spectropolarimetric data and
by the distribution of the core dominance parameter of BLRGs, HEGs and
LEGs (Chiaberge et al. 2002). This indicates that sources for which
line information are available (i.e. more than half of HEGs) are not
intrinsically less luminous.  Although these arguments certainly
cannot rule out a luminosity dependence of obscuration, we notice that
in any case the expected average luminosity correction between radio
galaxies and RLQs -- to account for the selection effects induced by
the 'receding torus' -- is typically estimated of order $\sim$ few
(compatible with the observed ratio in line luminosity of the two
populations, see \citealt{simp03}; \citealt{tadhu}). We conclude that
such an effect could not be the primary cause of the found bimodality
nor can significantly affect our results.

To summarize, selection effects should not qualitatively change the
observed distributions, and in particular the lack of sources at
intermediate $L_{\rm bol}$.

\subsection{Bolometric corrections and $L_{\rm bol}$}

We consider the assumptions made in deriving $L_{\rm bol}$ next.  The
value of the bolometric correction, $\eta$, adopted relies on an
average spectral energy distribution for RLQs.  On the basis of the
unification model, it is plausible to assume that this bolometric
correction is appropriate for BLRGs and HEGs as well, which are
believed to be intrinsically identical to RLQs.  However, while there
are indirect indications for low accretion luminosities in low power
RGs and BL Lac objects (from the lack of strong emission lines in
these systems, from limits on the amount of nuclear photon densities
and from the detected level of X-ray emission, e.g.
\citealt{celotti98}; \citealt{dimatteo00}), there is poor
observational and theoretical information on their spectral energy
distribution and even on the nature of the emission of the accretion
component. Therefore it is not at all obvious that a bolometric
correction which is valid for RLQs can also be applied to FR~I and low
excitation FR~II radio galaxies.

However, there are two additional ways of estimating $L_{\rm bol}$. 

The bolometric `disc' luminosity can be estimated from the observed
line luminosity, assumed to originate from gas photoionized by the
accretion radiation field.  This approach has several drawbacks,
including the fact that both the shape of the underlying photoionizing
continuum, and the covering factor of the line emitting gas are
unknown in these systems.  A comparison between $L_{\rm bol}$
estimated via a bolometric correction, and through the observed line
luminosity can be carried out for 49 radio galaxies with accurate and
homogeneous (in quality and available lines) [OII] and/or [OIII] line
luminosity information (see the on-line database maintained by Chris
Willott, http://www-astro.physics.ox.ac.uk/cjw/3crr /3crr.html; and
also \citealt{crawford88}; \citealt{rawlings89}; \citealt{cambia1};
\citealt{gelderman94}).  Assuming that the line radiation follows gas
photoionization by the nuclear continuum, from the line luminosities
we estimated the total luminosity in narrow lines (as $L_{\rm NLR}=
[3(3L_{\rm [OII]}+1.5L_{\rm [OIII]})]$, e.g.  \citealt{cambia1}), and
in turn the ionizing luminosity $L_{\rm ion} = L_{\rm NLR} C^{-1}$,
where $C$ is the covering factor. For a typical $C \sim 0.01$ for all
the LEGs (with the exclusion of 3C 123, which could be a
mis-classified HEG, Chiaberge et al. 2002) and the BLRGs we find a
good agreement - within a factor of a few - between the estimates of
$L_{\rm ion}$ and $L_{\rm bol}$ (inferred from the assumed
$\eta$). Consistently with the discussion above, HEGs have instead
$L_{\rm ion}$ comparable with that of BLRGs (and thus largely in
excess of their $L_{\rm bol}$). Seven (out of eleven) FR~Is show also
good agreement between the two estimates, while the remaining four
reveal an excess of $L_{\rm ion}$ by a factor 30-300.  In order to
further assess this issue, we also checked (via modeling using the
code CLOUDY, \citealt{ferland96}) on the possible difference in the
narrow line ratios (oxygen lines vs total luminosity) which could
arise if the (unknown) spectrum of FR~Is was significantly different
from the typical quasar-like spectrum.  We find that even in the cases
of spectra as hard as $F(\nu)\propto \nu^0$ the above estimates of
$L_{\rm ion}$ are robust (within factors of order unity). Significant
deviations are only achieved for harder spectral shapes or values of
the ionization parameter so low ($U \la$-4.5) to be insufficient to
give raise to the observed [OIII] luminosities. While the reason for
the observed discrepancy for the four FR~Is is not clear at this time,
we notice that for all four objects $L_{\rm NLR}$ has been derived
from the [OII] line luminosity only (by assuming a ratio of
[OII]/[OIII] fluxes, see \citealt{cambia1}) which could be more
significantly contaminated by emission not associated to nuclear
photoionization (such as shocks induced by the jets,
e.g. \citealt{dopita95}, or the line contribution from the galaxy
itself, see \citealt{baum95}). Note that the similarity of the
luminosities inferred by the two approaches in low power objects also
suggest that (at least for those sources) the levels of non-thermal
and 'disc' emission are in fact of the same order, reinforcing the
presence of low luminosity/low $\dot m$ peaks rather than tails in
these distributions.

For ten sources (9 RLQs and 1 BLRGs) we also estimated $L_{\rm ion}$
from the luminosity in broad lines (following the procedure in
\citealt{celpadghi97}; line luminosities from \citealt{jackson91})
finding again agreement with $L_{\rm bol}$ within a factor 1.5. Given
the uncertainties inherent to both methods, we conclude that there is
a good agreement between the luminosities $L_{\rm bol}$ estimated via
the bolometric correction $\eta$, and those estimated via the line
emission.

We should, however, point out that estimates of nuclear luminosities
based on line emission for a sample of BL~Lac objects (\citealt{wang02})
- which have typical extended radio luminosities similar to our sample
of RGs - are typically a factor 30 larger than those of our RGs based
on the assumption of a `universal' bolometric correction. This
discrepancy is possibly due to the fact that while \cite{wang02} used
both narrow and broad emission lines to estimate the bolometric
luminosity, they adopted a covering factor typical for broad line
emission ($C=0.1$).

It is also relevant to notice that no clear gap has been found in the
line emission properties of radio galaxies (e.g. \citealt{zirbaum95};
\citealt{baum95}; \citealt{willott99}) where the low frequency radio
emission has been found to correlate with the narrow line luminosity.
Nevertheless the results by the above authors clearly show the
presence of a transition (change of slope in such correlation) between
FR~Is and FR~IIs at values comparable to those inferred here, which
indicate that the line luminosity (for the same radio one) in FR~IIs is
larger by a factor $>$ 10 respect to what observed in FR~Is. In fact
\citet{willott00} discuss the possibility that an analysis including
the spectroscopic information on the line excitation might lead to a
clearer separation. Further support to this comes from the recent
work by \citet{wills04}, who targeted a sample of 13 low-luminosity RGs. 
The correlation between narrow line [OIII] luminosity and radio power 
in FR~II galaxies has the same slope, but significantly different zero 
point, than the one observed for FR~I galaxies. In particular, the 
latter have systematically lower optical line luminosity at a given 
radio power, in agreement with the results obtained by \citet{zirbaum95} 
and Baum et al. (1995).

The second, unfortunately even more uncertain, way of estimating
$L_{\rm bol}$ is through theoretical arguments.  If the accretion
regime in FR~Is and LEGs corresponds to an optically thick,
geometrically thin, radiatively efficient flow, the expected spectral
energy distribution could be approximated at first order as a
multicolor blackbody, plausibly similar in shape to that of RLQs, but
with a temperature and flux scaled to reflect the much lower mass
accretion rates.  In this case, the appropriate bolometric correction
would be a factor $\sim 4$ lower than the value derived for RLQs, in
the sense of moving the lower end of the distribution to lower
bolometric luminosities, thus making the separation between
non-thermal and thermal sources even more pronounced.  However, as
amply discussed in the literature, at low accretion rates the
accreting flow could attain a two temperature, optically thin,
geometrically thick flow structure. Although the dominant physical
processes, and the resulting conditions of the flow in such a regime
are still unclear, emission due to cyclo-synchrotron, Compton and
bremsstrahlung emission by a thermal electron population could be
expected (e.g.  \citealt{narayan97}). For instance, the most developed
spectral model \citep{quataert99} predicts a bolometric correction
which is about a factor $\sim$ 6 larger than the one used in our
analysis (for a typical set of parameters and $\dot m \sim 7\times
10^{-5}$) due to the contribution of the low energy cyclo-synchrotron
component.  This does work in the sense of `filling the gap' between
non-thermal and thermal sources, i.e. partly washing out the
bimodality in $L_{\rm bol}$.  [Note that the (partial) agreement of
the luminosities estimated via line emission and $L_{\rm bol}$ does
not affect this possibility, as the major component of the luminosity
in this case would not contribute to the gas photoionization].  We
however stress that this possibility inherently assumes that the
apparent bimodality arises as a consequence of a change in the flow
radiative efficiency. Indeed, as we will discuss in \S 5.2 the reduced
efficiency associated with an optically thin flow obviously acts in
the sense of `closing the gap' in the $\dot m$ as well as in the
$L_{\rm bol}$ distributions.

We conclude that in terms of their bolometric luminosity, radio-loud
sources are naturally divided in BLRG/RLQ, which show a relatively
broad peaked distribution at the high $L_{\rm bol}$ end, and FR~Is
plus low-excitation FR~IIs, populating the low $L_{\rm bol}$ peak.
Selection effects do not seem to account for the deficit of sources at
intermediate $L_{\rm bol}$. The observed distribution can be
reconciled with an underlying unimodal distribution apparently only if
the bolometric correction for the low $L_{\rm bol}$ sources has been
underestimated by a factor of several, as could be the case if e.g.
accretion in these sources proceeds through a geometrically thick,
optically thin accretion flow.

The fact that the BH mass does not play a role in determining the
different nuclear optical properties of the radio-loud populations,
the observation that the distribution in $\dot{m}$ show signs of being
more strongly bimodal than the one in $L_{\rm bol}$, as well as the
fact that $\dot m$ is ultimately regulating the emitted luminosity,
suggest that $\dot{m}$ can be the key parameter responsible for the
systematic difference in the emission of thermal and non-thermal
sources.

\subsection{Radiative efficiency vs $\dot m$}

We now discuss the role of the mass accretion rate in driving the
bimodal behavior of the different populations.  It is important to
stress that the sources with $L_{\rm bol}\la 3\times 10^{44}$ erg
s$^{-1}$ -- i.e. in between the two peaks of the $\dot m$ distribution
-- are HEGs, i.e.  objects partly absorbed, which should be
intrinsically as luminous as BLRGs.  This would imply the presence of
an underlying gap in the distribution spanning about one/two decades
in luminosity (see Fig.~\ref{fig2}, lower panel) and in accretion rate
(few $10^{-4}\la \dot m \la$ few$ 10^{-2}$ -- see the lower panel of
Fig.~\ref{fig3} in which HEGs are excluded). 
The presence of such a wide separation between the two peaks would
thus require a physical process able to account not only for a
transition, but also for a significant gap in the observed quantities.

A first possibility is that $\dot m$ is actually discontinuous, in the
sense that for $\dot m$ below a certain value ($\dot m \sim 0.01$) the
evolution time-scale of accretion is much faster than for higher $\dot
m$. This would account for a fast decrease in the number density of
'thermally' dominated sources and a peak/tail at much lower $\dot m$.
This could correspond to an on-off switching of the nuclear accretion
between different regimes in $\dot m$ (e.g. also \citealt{dimatteo03})
or an evolutionary sequence from the FR~II to the FR~I-like regime
(\citealt{zirbaum95}; \citealt{ledlow96};
\citealt{ghisellinicelotti2001}; \citealt{cavalelia}).

Alternative possibilities where all $\dot{m}$ are possible, equally
probable and have a continuous distribution are however open. The
transition could then be due to a change (but {\it not} simply a
smooth decline as predicted by simple advection flow models) in the
radiative efficiency $\epsilon$ of the accreting flow.  Theoretically,
the transition between a radiative efficient, geometrically thin,
optically thick accretion flow, and a radiatively inefficient,
geometrically thick and optically thin configuration is expected to
occur at $\dot{m} \sim \alpha^2$ (\citealt{rees82};
\citealt{narayan95}). For a viscosity parameter $\alpha\sim 0.1$,
considered to be phenomenologically and physically plausible (e.g.
\citealt{mahadevan97}; \citealt{balbushawley98}), this value is indeed
close to the transition in the $\dot{m}$ distribution shown in
Fig.~\ref{fig3}: in this scenario the peak at low $\dot{m}$ is an
artificial consequence of our assumption that $\epsilon$ is the same
for all sources, while in fact the radiative efficiency for the
objects at low $\dot{m}$ is much smaller (and possibly a decreasing
function of $\dot{m}$ itself, \citealt{narayan95}).  However our
limited sample and the theoretical uncertainties do not allow to
expand further on this point, nor (yet) to give any robust indication
on the value of $\alpha$. A similar interpretation has been
suggested by \citet{wills04} in relation to the offset in the [OIII]
luminosity between FR~Is and FR~IIs. However, another plausible
possibility in a two temperature optically thin configuration is that
the flow is subject to significant energy/mass loss, in the form of an
outflow, as pointed out by \citet{blandbeg99} (see also
\citealt{hawleybalbus02}; and for a different view
\citealt{narayan97c}; \citealt{narayan00}) or a jet (e.g.
\citealt{pellegrini03}). In such a case in fact the actual $\dot m$
accreting onto the BH would be much smaller than that at large disc
radii. Note also that a bimodal distribution in $\dot{m}$, and
especially the presence of a gap, might conflict with the proposed
possibility of an increase -- if continuous -- of magnetic dissipation
of disc energy into the corona (possibly associated with the formation
of jet) at decreasing mass accretion rates (\citealt{merloni02};
\citealt{livio03}). Clearly in these situations, as discussed in \S
5.1, also the intrinsic distribution in $L_{\rm bol}$ can of course be
continuous.

Different scenarios can also be envisioned, as at values of $\dot m
\sim 10^{-2}$ other physical processes could be relevant. For
instance, in the optically thick, high surface density disc solutions
a transition from the gas pressure to radiation pressure dominated
accretion regime for supermassive BH can occur, for `reasonable'
values of the viscosity parameter: this could thus imply the presence
of an unstable flow in the inner disc onsetting at these values of
$\dot m$ \citep{shakura73}.  While a paucity of sources over a narrow
range in $\dot m$ would disfavor such possibility, an actual gap could
in fact be consistent with this view.

Alternatively, the intermediate region could correspond to a range of
$\dot m$ for which the formation of a powerful jet is inhibited. In
MHD models of relativistic jet formation, and in particular in the
\citet{blandzna77} one in which the black hole rotational energy is
extracted via a Penrose-like process, the key parameter determining
the efficiency of the process is the strength of the poloidal magnetic
field $B_{\rm p}$. If, as stressed by \citet{livio99}, a realistic
condition for the equilibrium poloidal magnetic field is $B_{\rm p}
\sim (H/R) B_{\rm \phi}$ (where $H/R$ is the inner height to distance
ratio in the disc and $B_{\rm \phi}$ is the azimuthal component of the
magnetic field), then the process of jet formation could be efficient
enough only for $H/R \sim$1, i.e. either for $\dot{m} \ll$1
(accretion-starved advective disc) or for $\dot{m} \ga$1
(super-Eddington advective-disc), as suggested by \citet{meier02}.
Again, while the presence of a significant number of sources at
intermediate $\dot m$ may cast doubts on such interpretation, the
assumption that HEGs are intrinsically identical to BLRGs, leads to an
absence of sources in a range of almost two decades in $\dot m$.

As a final note, we point out that for $\epsilon \sim 0.1$ some RLQs
result to be super-Eddington, although by only a factor of a few.  In
order to ascertain whether these RLQs could be partly affected by
relativistic beaming, we estimated their radio spectral indices, and
found that only five (out of the 17 RLQs above the Eddington value)
are actually flat spectrum sources, indicating that relativistic
beaming likely does not largely account for the highest
luminosities. Also, the majority of these sources are not the highest
redshift RLQs in our sample (six are at $z<$0.3 and only five are at
$z>$0.8), and thus evolutionary effects do not largely account for
their extreme luminosities.

\section{Conclusions}

We considered the nuclear properties of radio-loud objects, comprising
a complete sample of radio galaxies (both FR~I and FR~II) and
radio-loud quasars. We estimated the mass of the central supermassive
black hole, the nuclear accretion luminosities, and the corresponding
mass accretion rates.

We found that while BH masses span a relatively large range, they are
clustered around $10^8-10^9$ M$_{\odot}$ and there is no systematic
difference in their values between FR~I, FR~II radio galaxies and
radio-loud quasars, in agreement with previous findings
(e.g. \citealt{odowd02}). Biases introduced by the (radio) criteria
adopted in defining the samples do not allow us to address whether a
lower limit on the BH mass or an upper envelope in radio luminosity
for a given mass exist in radio-loud sources
(\citealt{McLureDunlop2002}; \citealt{laor00}; \citealt{dunl03}; but
see \citealt{ho2002}; \citealt{oshlack2002}; Woo \& Urry 2002a,
2002b).

However, the distributions in bolometric luminosity, and (even more
so) in the corresponding mass accretion rate appear to be
bimodal. FR~I, low excitation FR~II and some high excitation FR~II
galaxies occupy the low end of the $L_{\rm bol}$ and $\dot{m}$
distributions: $10^{41}\la L_{bol} \la 10^{43.5}$ erg s$^{-1}$ or
$10^{-6}\la \dot{m} \la$ few $10^{-4}$. On the other hand, radio-loud
quasars, broad line radio galaxies and some high excitation FR~II
galaxies occupy the high end of the distributions: $10^{45}\la L_{\rm
bol} \la 10^{47}$ erg s$^{-1}$ or $10^{-2}\la \dot{m} \la$ 1. The
region in between is characterized by a marked deficiency of sources,
especially if high excitation galaxies are indeed intrinsically
similar to broad line ones.  The relative completeness and homogeneity
of our sample, which includes low and high power sources selected on
the basis of the extended radio properties, allowed to reveal such
bimodal behavior, which had not been noticed in previous studies
(\citealt{odowd02}; \citealt{woo02a}). This is so pronounced to appear
despite of the uncertainties in the inferred quantities. No clear
selection bias which could induce such result has been found. In a
future work, we will test this bimodal behavior through more accurate
determination of the BH masses (via the $M_{\rm BH}-\sigma$ relation)
and luminosities for a smaller subsample of sources including a
representative sample of FR~II sources (Marchesini, Ferrarese \&
Celotti, in preparation). The range of $\dot m$ corresponding to the
transition/gap between FR~Is and FR~IIs ($\dot{m}\sim 10^{-2}-10^{-1}$ for
$\epsilon =0.1$) is similar to the value of $\dot m \sim$0.06
corresponding to the dividing line between FR~I and FR~II radio
galaxies discussed by \citet{ghisellinicelotti2001}, which was
inferred via an independent procedure based on the properties of the
\citet{ledlow96} RG sample via global correlations between the
extended radio emission and the accretion one. In particular the
magnitudes of the host galaxies reported by \citet{ledlow96} were
converted into BH masses (again via the $M_{\rm bulge} - M_{\rm BH}$
relation), while the radio luminosities were associated to nuclear
accretion rates, by adopting the correlation between radio power and
narrow line luminosity \citep{willott99}, and the latter converted
into ionization luminosity. The absence of a gap in the findings by
\citet{ghisellinicelotti2001} might however arise as the relation by
\citet{willott99} was (necessarily) adopted by these authors also for
FR~Is, which tends in fact to overestimate the FR~I line luminosity
(as also shown by \citealt{wills04}).

A dichotomy is not seen if the radio luminosity is substituted to
$L_{\rm bol}$ (in this case the distribution appears continuous with
significant overlap between the different classes), making our
findings even more intriguing.  A discontinuity in the $\dot{m}$
distribution implies a transition/gap in the accretion which is not
reflected in the extended radio emission and possibly, if a similar
jet radiative efficiency would characterize FR~Is and FR~IIs, in the
jet power. This would also imply a significantly different ratio
between the accretion and jet powers in low and high luminosity
systems, questioning whether the $\dot m$ taken at face value in low
power sources is sufficient to give raise to the associated jets. We
also notice that the radio luminosity corresponding to the transition
in $L_{\rm opt}$ between LEGs and BLRGs -- according to the
established radio-line correlations (e.g. \citealt{willott99}) --
reflect the break in the radio sources luminosity function
(e.g. \citealt{dunlpeak}).

Our findings indicate that the whole radio-loud population can be
characterized by thermal vs non-thermal dominated (optical) emission
(as suggested by Chiaberge et al. 2002).  The population of FR~II
radio galaxies is in this respect complex and - as expected - the
source behavior is related to the nuclear line properties.  In other
words, the nuclear structure and the central activity define the
parent radio galaxies of RLQs and BL~Lacs and such activity is not
uniquely connected to the extended radio power and morphology. This
might suggest different time-scales for the evolution of the radio and
optical activity (see \citealt{ciras03}) and/or a temporal evolution
from the higher to the lower power sources.

Some possibilities accounting for the found bimodal behavior have been
discussed.  These involve a relatively rapid transition in the
accretion rate; a transition in the accretion mode, implying a change
in radiative efficiency and/or the onset of significant mass loss,
with transition values of $\dot{m}$ in agreement with the rates at
which radiative inefficient flows are expected to be possible; the
presence of an unstable (e.g.  radiation pressure dominated) accretion
regime or an inefficient jet formation process associated to
geometrically thin flows at intermediate $\dot m$. While no definite
conclusion can be drawn on the origin of the found transition/gap,
some of these hypothesis might be (dis)-favored by examining with more
accurate measures the effective absence of sources over a significant
range at intermediate $\dot m$, i.e. the presence of a significant gap
in the observed properties.  A further aspect which we are exploring
is whether the dichotomic behavior holds also for radio-quiet sources,
which would imply that the last of the above possibilities cannot be
the (only) explanation.

Several pieces of evidence already indicated that in low power radio
galaxies and BL~Lac objects the luminosity resulting from dissipation
in the accretion flow is extremely low, compared to that typically
inferred from RLQ activity (as e.g. from the low line luminosity,
limits on the nuclear emission, e.g. \citealt{urry95};
\citealt{celotti98}; \citealt{odowd02}; \citealt{wang02}).
Furthermore, stringent limits came with the finding that the radiated
output is orders of magnitude lower than that expected from the
(Bondi) accretion rate estimated from the properties of hot gas at the
accretion radius for a 10 per cent efficiency (\citealt{fabian95};
e.g. M87, \citealt{dimatteo03}; IC4296, \citealt{pellegrini03}).  Our
findings, consistent with the above results, appear thus to indicate
the presence of a relatively rapid transition and a gap between the
high radiative efficient accretion regime and the accretion properties
of low power sources.  Clearly, these results are also consistent with
the possibility, suggested by \citet{dimatteo03}, that the low disc
accretion rate might have a time dependent (on-off) character (even
between low activity and almost `inactive' states).  Note that if
indeed a large fraction of the radio, X--ray and optical emission in
some low power objects is dominated by non-thermal radiation from the
base of the jet (as proposed), this might imply $\dot{m}$ lower than
$\sim 10^{-5}-10^{-6}$. The findings from the line emission however
indicate that at least in a relevant number of sources the non-thermal
jet and 'disc' emission should be of similar order.

Finally, it is worth mentioning that the found behavior could indeed
be analogous to the transitions between spectral states present in
black hole X-ray binary systems and microquasars, possibly in terms of
the value of the transitional $\dot m$, the nuclear spectral
properties and the onset and nature of the jet production
(e.g. \citealt{fender}).

As already amply stressed in the literature in relation to the
understanding of accretion at low rates, there is great need to test
for the presence of reflection features in the X-ray spectra of such
systems, as well as variability and polarization measures, as it has
already been possible for the Galactic Center (\citealt{baganoff01};
\citealt{aitken00}). Theoretical and numerical work might however lead
to a deeper understanding first.

\section*{Acknowledgments}
We would like to thank Arunav Kundu for making available the code KMM
from \citet{kmm94}. We are also grateful to Gary Ferland for
maintaining his freely distributed code, CLOUDY. We also acknowledge
the anonymous referees for constructive criticism which helped
improving the paper. This research has made use of the NASA/IPAC
Extragalactic Database (NED) which is operated by the Jet Propulsion
Laboratory, California Institute of Technology, under contract with
the National Aeronautics and Space Administration.  DM and AC
acknowledge the Italian MIUR and INAF for financial support. LF
acknowledges support provided by NASA through LTSA grant number
NAG5-8693.

\bsp

\newpage

\begin{table}\label{tbl2}
  \caption{Properties of the sources}
  \begin{tabular}{@{}lccr}
  \hline
Name    &  $M_{B}$  &  $\log{M_{BH}}$   &  $\log{L_{bol}}$  \\
            &           &  ($M_{\odot}$)    &  (erg s$^{-1}$)   \\
 \hline
\multicolumn{4}{c}{3CR~FR~I} \\
 \hline
       3CR~28    &  -21.01  &     8.67 &    $<$42.61 \\
       3CR~29    &  -21.29  &     8.81 &    42.45 \\
       3CR~31    &  -19.40  &     7.89 &    41.99 \\
       3CR~66    &  -21.35  &     8.84 &    42.72 \\
       3CR~78    &  -21.63  &     8.98 &    43.67 \\
       3CR~83.1  &  -21.70  &     9.01 &    41.31 \\
       3CR~84    &  -22.26  &     9.28 &    44.03 \\
       3CR~89    &  -21.33  &     8.83 &    42.06 \\
       3CR~264   &  -21.38  &     8.85 &    43.14 \\
       3CR~270   &  -20.81  &     8.57 &    40.89 \\
       3CR~272.1 &  -20.35  &     8.35 &    41.36 \\
       3CR~274   &  -20.16  &     8.26 &    42.18 \\
       3CR~277.3 &  -20.66  &     8.50 &    42.47 \\
       3CR~288   &  -21.11  &     8.72 &    43.20 \\
       3CR~296   &  -21.94  &     9.13 &    41.65 \\
       3CR~310   &  -20.23  &     8.29 &    42.40 \\
       3CR~314.1 &  -20.59  &     8.47 &    $<$42.59 \\
       3CR~317   &  -21.27  &     8.80 &    42.52 \\
       3CR~338   &  -22.14  &     9.23 &    42.34 \\
       3CR~346   &  -21.46  &     8.89 &    44.28 \\
       3CR~348   &  -21.35  &     8.84 &    42.77 \\
       3CR~424   &  -19.97  &     8.17 &    42.85 \\
       3CR~438   &  -21.28  &     8.80 &    43.14 \\
       3CR~442   &  -20.21  &     8.28 &    41.16 \\
       3CR~449   &  -19.04  &     7.71 &    42.13 \\
       3CR~465   &  -22.33  &     9.32  &   42.61 \\
 \hline
\multicolumn{4}{c}{B2 radio-galaxies} \\
 \hline
       0055+26  &  -21.84 &     9.08  &    $<$41.84 \\
       0055+30  &  -22.06 &     9.18  &    42.40 \\
       0120+33  &  -21.56 &     8.94  &    $<$40.67 \\
       0708+32  &  -20.83 &     8.59  &    $<$42.34 \\
       0755+37  &  -21.54 &     8.93  &    43.22 \\
       0908+37  &  -21.81 &     9.06  &    43.14 \\
       0924+30  &  -20.85 &     8.60  &    $<$41.02 \\
       1003+26  &  -20.90 &     8.62  &    $<$41.99 \\
       1005+28  &  -21.58 &     8.95  &    $<$42.70 \\
       1113+24  &  -22.11 &     9.21  &    $<$41.86 \\
       1204+34  &  -20.82 &     8.58  &    $<$42.88 \\
       1217+29  &  -19.16 &     7.77  &    40.47 \\
       1257+28  &  -21.05 &     8.69  &    $<$40.52 \\
       1346+26  &  -21.73 &     9.02  &    42.64 \\
       1422+26  &  -20.45 &     8.40  &    $<$42.45 \\
       1430+25  &  -20.92 &     8.63  &    $<$42.68 \\
       1447+27  &  -20.35 &     8.35  &    $<$42.06 \\
       1450+28  &  -21.20 &     8.76  &    $<$42.88 \\
       1512+30  &  -21.51 &     8.92  &    $<$42.07 \\
       1521+28  &  -21.32 &     8.82  &    43.28 \\
       1527+30  &  -22.40 &     9.35  &    $<$42.99 \\
       1557+26  &  -21.04 &     8.69  &    $<$42.73 \\
       1610+29  &  -20.95 &     8.64  &    $<$41.42 \\
       1613+27  &  -21.33 &     8.83  &    $<$42.30 \\
       1658+30A &  -19.85 &     8.11  &    $<$42.61 \\
       1827+32  &  -21.53 &     8.93  &    $<$42.44 \\
       2116+26  &  -20.86 &     8.60  &    41.23 \\
       2236+35  &  -20.75 &     8.55  &    41.58 \\
 \hline
\end{tabular}
\end{table}

\begin{table}
  \begin{tabular}{@{}lccr}
  \hline
Name    &  $M_{B}$  &  $\log{M_{BH}}$   &  $\log{L_{bol}}$  \\
            &           &  ($M_{\odot}$)    &  (erg s$^{-1}$)   \\
 \hline
\multicolumn{4}{c}{3CR~FR~II} \\
 \hline
       3CR~15    &  -21.07  &     8.70  &   $<$42.75 \\
       3CR~17    &  -20.44  &     8.39  &   44.75 \\
       3CR~18    &  -21.52  &     8.92  &   44.45 \\
       3CR~33.1  &  -20.35  &     8.35  &   44.40 \\
       3CR~35    &  -21.19  &     8.76  &   $<$42.48 \\
       3CR~63    &  -20.81  &     8.58  &   44.12 \\
       3CR~79    &  -20.62  &     8.48  &   44.16 \\
       3CR~88    &  -20.71  &     8.53  &   42.43 \\
       3CR~98    &  -20.09  &     8.23  &   $<$41.50 \\
       3CR~111   &  -22.85  &     9.56  &   45.58 \\
       3CR~123   &  -18.51  &     7.45  &   $<$42.62 \\
       3CR~132   &  -22.03  &     9.17  &   $<$43.12 \\
       3CR~133   &  -22.74  &     9.53  &   44.76 \\
       3CR~135   &  -20.56  &     8.45  &   43.20 \\
       3CR~153   &  -21.88  &     9.10  &   $<$42.71 \\
       3CR~165   &  -23.49  &     9.88  &   43.12 \\
       3CR~166   &  -20.58  &     8.46  &   44.05 \\
       3CR~171   &  -19.55  &     7.96  &   42.33 \\
       3CR~184.1 &  -20.30  &     8.32 &    44.00 \\
       3CR~192   &  -20.07  &     8.21  &   $<$42.89 \\
       3CR~197.1 &  -21.48  &     8.90  &   43.96 \\
       3CR~198   &  -19.82  &     8.09  &   43.86 \\
       3CR~219   &  -21.27  &     8.80  &   44.47 \\
       3CR~223   &  -20.66  &     8.50  &   $<$43.10 \\
       3CR~223.1 &  -20.77  &     8.56  &   $<$43.01 \\
       3CR~227   &  -20.31  &     8.33  &   44.68 \\
       3CR~234   &  -21.67  &     8.99   &  44.94 \\
       3CR~236   &  -20.91  &     8.62   &  $<$42.86 \\
       3CR~285   &  -20.58  &     8.46   &  41.48 \\
       3CR~287.1 &  -20.73  &     8.54   &  44.54 \\
       3CR~300   &  -19.91  &     8.14   &  43.62 \\
       3CR~303   &  -20.51  &     8.43   &  44.50 \\
       3CR~318.1 &  -19.69  &     8.03   &  $<$41.51 \\
       3CR~319   &  -19.80  &     8.08   &  $<$43.24 \\
       3CR~323.1 &  -22.99  &     9.64   &  46.01 \\
       3CR~326   &  -19.91  &     8.13  &   $<$42.95 \\
       3CR~327   &  -21.67  &     9.00  &   $<$42.38 \\
       3CR~332   &  -21.03  &     8.68  &   44.62 \\
       3CR~349   &  -20.10  &     8.23  &   43.99 \\
       3CR~353   &  -18.79  &     7.59  &   $<$41.37 \\
       3CR~357   &  -21.89  &     9.10  &   $<$42.73 \\
       3CR~379.1 &  -21.65  &     8.98  &   $<$43.10 \\
       3CR~381   &  -20.90  &     8.62  &   $<$43.35 \\
       3CR~382   &  -21.80  &     9.06  &   45.55 \\
       3CR~388   &  -22.18  &     9.24  &   43.09 \\
       3CR~390.3 &  -20.48  &     8.41  &   44.87 \\
       3CR~401   &  -19.95  &     8.16  &   43.51 \\
       3CR~402   &  -20.01  &     8.18  &   42.43 \\
       3CR~403   &  -20.48  &     8.41  &   42.49 \\
       3CR~445   &  -20.31  &     8.33   &  45.10 \\
       3CR~452   &  -20.75  &     8.54   &  $<$42.73 \\
       3CR~456   &  -19.21  &     7.80   &  44.33 \\
       3CR~460   &  -20.91  &     8.63   &  43.44 \\
 \hline
\end{tabular}
\end{table}

\begin{table}
  \begin{tabular}{@{}lccr}
  \hline
Name    &  $M_{B}$  &  $\log{M_{BH}}$   &  $\log{L_{bol}}$  \\
            &           &  ($M_{\odot}$)    &  (erg s$^{-1}$)   \\
 \hline
\multicolumn{4}{c}{3CR~QSO} \\
 \hline
       3CR~14    &  -22.61  &    9.45  &   46.29 \\
       3CR~43    &  -21.13  &    8.73  &   45.54 \\
       3CR~47    &  -20.71  &    8.52  &   45.56 \\
       3CR~93    &  -20.37  &    8.36  &   45.10 \\
       3CR~138   &  -21.12  &    8.73  &   45.87 \\
       3CR~147   &  -21.71  &    9.02  &   45.99 \\
       3CR~154   &  -21.24 &     8.79  &   46.30 \\
       3CR~175   &  -22.58  &    9.44  &   46.77 \\
       3CR~179   &  -21.45  &    8.89  &   45.91 \\
       3CR~186   &  -22.82  &    9.56  &   46.58 \\
       3CR~190   &  -21.50  &    8.91  &   46.14 \\
       3CR~191   &  -24.31  &   10.29  &   46.99 \\
       3CR~204   &  -21.97  &    9.14  &   46.58 \\
       3CR~207   &  -21.75  &    9.03  &   46.13 \\
       3CR~215   &  -20.69  &    8.52  &   45.52 \\
       3CR~216   &  -21.10  &    8.72  &   45.42 \\
       3CR~249.1 &  -22.67  &    9.48  &   45.68 \\
       3CR~254   &  -21.96   &   9.14  &   46.26 \\
       3CR~263   &  -21.99  &    9.15  &   46.84 \\
       3CR~268.4 &  -23.55  &    9.91  &   46.92 \\
       3CR~270.1 &  -23.42  &    9.85  &   46.67 \\
       3CR~277.1 &  -20.64  &    8.49  &   45.20 \\
       3CR~280.1 &  -23.49  &    9.88  &   46.55 \\
       3CR~298   &  -24.05  &   10.15  &   47.36 \\
       3CR~309.1 &  -22.82  &    9.56  &   46.57 \\
       3CR~334   &  -21.88  &    9.10  &   46.44 \\
       3CR~380   &  -22.44  &    9.37  &   46.37 \\
       3CR~418   &  -22.38  &    9.32  &   46.07 \\
       3CR~432   &  -24.08  &   10.17  &   46.93 \\
       3CR~454   &  -23.83   &  10.05  &   46.54 \\
       3CR~455   &  -20.73   &   8.53  &   45.05 \\
 \hline
\multicolumn{4}{c}{RLQ from McLure \& Dunlop (2001)} \\
 \hline
       0137+012   &   -21.59  &    8.96  &   45.31 \\
       0736+017   &   -21.13  &    8.73  &   45.67 \\
       1004+130   &   -21.65  &    8.99  &   46.29 \\
       1020-103   &   -20.91  &    8.62  &   44.95 \\
       1217+023   &   -21.26  &    8.80  &   45.93 \\
       1226+023   &   -21.95  &    9.13  &   46.33 \\
       1302-102   &   -21.05  &    8.69  &   45.93 \\
       2135-147   &   -20.92  &    8.63  &   46.24 \\
       2141+175   &   -20.98  &    8.66  &   46.22 \\
       2247+140   &   -21.35  &    8.84  &   45.64 \\
       2349-014   &   -21.87  &    9.09  &   46.01 \\
       2355-082   &   -21.17  &    8.75  &   45.07 \\
 \hline
\multicolumn{4}{c}{RLQ from Hutchings et al.(2002)} \\
 \hline
       0033+098  &    -23.00  &    9.64 &  46.72 \\
       0225-014  &    -22.00  &    9.16 &  46.88 \\
       0820+296  &    -22.71  &    9.50 &  46.74 \\
 \hline
\end{tabular}
\end{table}

\begin{table}\label{tbl1}
  \caption{Results of statistical tests}
  \begin{tabular}{@{}llrrrr}
  \hline
        &         & \multicolumn{3}{c}{Kolmogorov-Smirnov Test $p$}  \\
Pop. 1  & Pop. 2  &       $M_{\rm BH}$          & $L_{\rm bol}$                 
&  $\dot{m}$            \\
 \hline

FR~I     & FR~II    & $\dots$  & 7$\times$10$^{-8}$  & 2$\times$10$^{-7}$  \\
FR~I     & RLQ     & 2$\times$10$^{-3}$ & 3$\times$10$^{-22}$ & 3$\times$10$^{-22}$ \\
FR~II    & RLQ     & 5$\times$10$^{-7}$ & 5$\times$10$^{-20}$ & 1$\times$10$^{-17}$ \\
3CR FR~I & B2RG    & $\dots$            & $\dots$             & $\dots$             \\
LEG     & HEG     & $\dots$            & $\dots$             & $\dots$             \\
LEG     & BLRG    & $\dots$            & 7$\times$10$^{-6}$  & 7$\times$10$^{-6}$  \\
HEG     & BLRG    & $\dots$            & 5$\times$10$^{-8}$  & 7$\times$10$^{-7}$  \\
3CR RLQ   & MDRLQ   & $\dots$            & $\dots$             & $\dots$             \\
FR~I     & LEG     & $\dots$            & $\dots$             & $\dots$             \\
FR~I     & HEG     & $\dots$            & 1$\times$10$^{-4}$  & 3$\times$10$^{-5}$  \\
FR~I     & BLRG    & $\dots$            & 5$\times$10$^{-10}$  & 5$\times$10$^{-10}$ \\
3CR FR~I & 3CR FR~II& $\dots$            & 3$\times$10$^{-4}$  & 1$\times$10$^{-4}$  \\
  \hline
\multicolumn{2}{c}{}            & \multicolumn{3}{c}{Means and standard 
deviations}       \\
\multicolumn{2}{c}{Population}  & $\log{M_{\rm BH}}$ & $\log{L_{\rm bol}}$ & $\log{\dot{m}}$    \\
 \hline
\multicolumn{2}{c}{FR~I}   & 8.7$\pm$0.4   & 42.30$\pm$0.83  & -4.51$\pm$0.79     \\
\multicolumn{2}{c}{B2RG}   & 8.8$\pm$0.3   & 42.16$\pm$0.81  & -4.72$\pm$0.75     \\
\multicolumn{2}{c}{FR~II}  & 8.6$\pm$0.5   & 43.53$\pm$1.09  & -3.14$\pm$1.03     \\
\multicolumn{2}{c}{LEG}    & 8.5$\pm$0.7   & 42.70$\pm$0.64  & -3.89$\pm$0.69     \\
\multicolumn{2}{c}{HEG}    & 8.5$\pm$0.4   & 43.19$\pm$0.82  & -3.41$\pm$0.91     \\
\multicolumn{2}{c}{BLRG}   & 8.7$\pm$0.5   & 44.92$\pm$0.50  & -1.91$\pm$0.30     \\
\multicolumn{2}{c}{RLQ}    & 9.1$\pm$0.5   & 46.13$\pm$0.60  & -1.11$\pm$0.36     \\
 \hline
        &         & \multicolumn{3}{c}{KKM Test}  \\
Parameter       & $\mu_{1}$ & $\mu_{2}$ & $\sigma^{2}$ & P-value \\
\hline
                & \multicolumn{4}{c}{All radio-loud sources}    \\
$\log{L_{\rm bol}}$  & 42.60 & 45.77 & 0.76 & $<$0.001 \\
$\log{\dot{m}}$      & -4.30 & -1.49 & 0.54 & $<$0.001 \\
                & \multicolumn{4}{c}{3CR radio-loud sources}    \\
$\log{L_{\rm bol}}$  & 42.85 & 45.76 & 0.79 & $<$0.001 \\
$\log{\dot{m}}$      & -4.09 & -1.60 & 0.49 & $<$0.001 \\
                & \multicolumn{4}{c}{3CR RGs}        \\
$\log{L_{\rm bol}}$  & 42.58 & 44.42 & 0.51 & 0.101 \\
$\log{\dot{m}}$      & -4.20 & -2.39 & 0.43 & 0.032 \\
 \hline
\end{tabular}

\medskip $M_{\rm BH}$ and $L_{\rm bol}$ are in units of $M_{\odot}$ and
erg~s$^{-1}$, respectively. According to the KMM algorithm \citep{kmm94}
$\mu_{1}$ and $\mu_{2}$ are the means of the two Gaussian distributions
representing the bimodality, while $\sigma^{2}$ is their common covariance;
when the P-value is $<$0.05, the observed distribution is strongly inconsistent
with being unimodal; when P-value $\in$ [0.05,0.1] the observed
distribution is only marginally inconsistent with being unimodal;
therefore, when P-value is $>$0.1, the observed distribution is not
inconsistent with unimodality. For the Kolmogorov-Smirnov statistical test,
only values for the significantly different populations are reported.
\end{table}


\end{document}